\pdfoutput=1

\documentclass[11pt]{article}

\usepackage{ACL2023}

\usepackage{times}
\usepackage{latexsym}
\usepackage{multicol}
\usepackage{booktabs}
\usepackage[fleqn]{amsmath}
\usepackage{multirow}
\usepackage[normalem]{ulem}
\usepackage{hyperref}
\usepackage{graphicx}
\usepackage{subfigure}
\usepackage{fixltx2e}
\usepackage{algorithm}
\usepackage{algorithmic}
\usepackage{setspace}
\usepackage{soul}
\usepackage{amsmath}
\usepackage{amsthm}

\usepackage[T1]{fontenc}
\usepackage[utf8]{inputenc}

\usepackage[utf8]{inputenc}

\usepackage{microtype}

\usepackage{inconsolata}

\title{CHIQ: Contextual History Enhancement for Improving Query Rewriting in Conversational Search}

\author{Fengran Mo$^1$, Abbas Ghaddar$^{2}$, Kelong Mao$^{3}$,\\ \textbf{Mehdi Rezagholizadeh$^{2}$, Boxing Chen$^{2}$, Qun Liu$^2$, Jian-Yun Nie$^{1}$}\\
$^1$DIRO, Université de Montréal, Québec, Canada \\
$^2$Huawei Noah’s Ark Lab, Montreal Research Center, Canada \\ 
$^3$Renmin University of China \\
\texttt{fengran.mo@umontreal.ca, nie@iro.umontreal.ca} \\
}

\usepackage{booktabs}
\usepackage{eqparbox}
\usepackage{arydshln}
\usepackage{subcaption}
\usepackage{rotating,csquotes}
\usepackage{xcolor}
\usepackage{multirow}
\usepackage{framed}
\usepackage{booktabs}
\usepackage{amsmath}
\usepackage{amssymb}
\usepackage{mathtools}
\usepackage{upgreek}
\usepackage{soul}
\usepackage{amsfonts}
\usepackage{todonotes}
\usepackage[most]{tcolorbox}
\usepackage{tabularx}

\newtcbox{\mybox}[1][]{enhanced, colframe=blue, colback=blue!15, 
	frame style={opacity=0.25}, interior style={opacity=0.25}, 
	nobeforeafter, tcbox raise base, shrink tight, extrude by=1mm, #1}

\newcommand{\fname}{\textsc{CHIQ}}
\usepackage{amssymb}
\usepackage{pifont}
\newcommand{\cmark}{\ding{51}}%
\newcommand{\xmark}{\ding{55}}%

\newcommand{\chatgpt}{\textsc{ChatGPT}}

\DeclareMathAlphabet\mathbfcal{OMS}{cmsy}{b}{n}

\newcommand{\mightmention}[1]{}
\newcommand{\problem}[1]{\textcolor{red}{$\star$}}
\newcommand{\answer}[1]{\textcolor{blue}{$\#$}}
\newcommand{\todoreview}[1]{\textcolor{green}{$@$}}

\usepackage[most]{tcolorbox}
\newcommand{\specialcell}[2][c]{%
  \begin{tabular}[#1]{@{}c@{}}#2\end{tabular}}

\begin{document}
\maketitle



\begin{abstract}
In this paper, we study how open-source large language models (LLMs) can be effectively deployed for improving query rewriting in conversational search, especially for ambiguous queries. We introduce \fname{}, a two-step method that leverages the capabilities of LLMs to resolve ambiguities in the conversation history before query rewriting. This approach contrasts with prior studies that predominantly use closed-source LLMs to directly generate search queries from conversation history. We demonstrate on five well-established benchmarks that \fname{} leads to state-of-the-art results across most settings, showing highly competitive performances with systems leveraging closed-source LLMs. Our study 
provides a first step towards leveraging open-source LLMs in conversational search, as a competitive alternative to the prevailing reliance on commercial LLMs for query rewriting. Our code is publicly available at \url{https://github.com/fengranMark/CHIQ}.

\end{abstract}

\begin{figure*}[!th]
    \centering
    \includegraphics[width=1\linewidth]{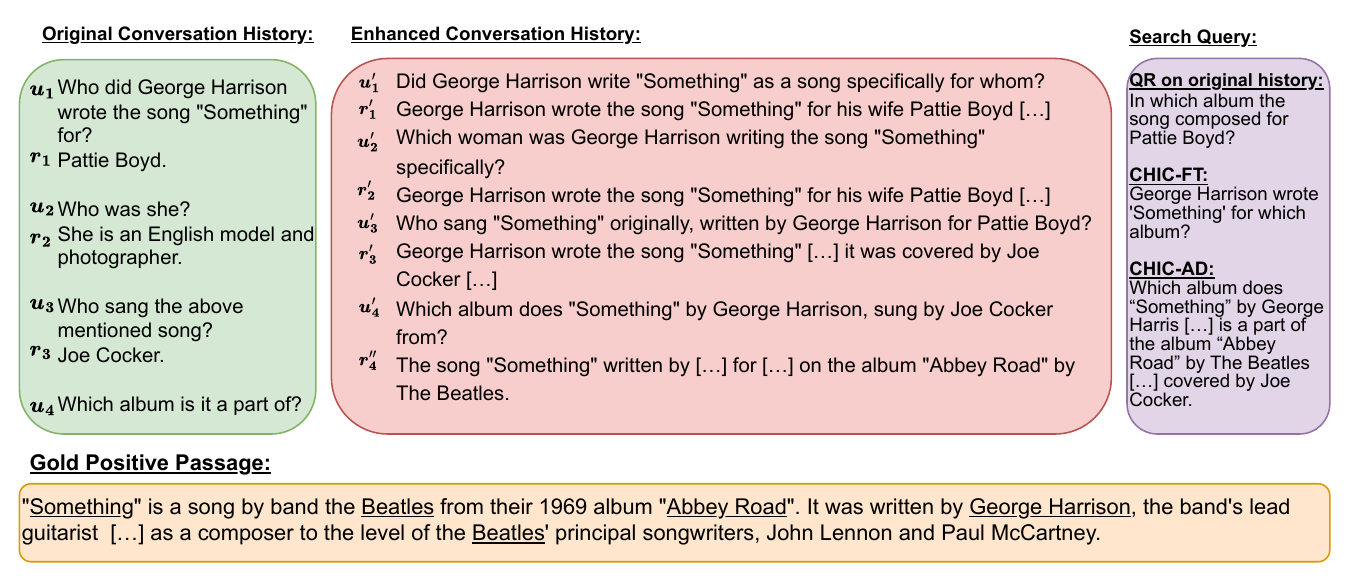}
    \caption{An illustrative example of a conversational history (left box) and the gold positive passage relevant to the last user turn. The enhanced history obtained using our method described in \S~\ref{sec:History Enhancement} is in the middle box. The right box shows the three search queries generated by LLaMA-2-7B conditioned on the original history, and our \fname{-FT} and \fname{-AD} methods described in~\S~\ref{sec:QR Finetuning} and ~\S~\ref{sec:Ad-hoc QR}, respectively. \underline{Underlined} terms in the gold passages are those that appear in the query generated by our approaches, which is conditioned on the enhanced history and did not appear in the query generated by the method that uses the original history.}
    \label{fig:intro}
\vspace{-2ex}
\end{figure*}

\section{Introduction}
Conversational search enables users to interact with the system in a multi-turn fashion to satisfy their complex information needs~\cite{gao2022neural,zamani2023conversational}. One of the crucial steps is to compose adequate search queries for each context-dependent utterance.
Recent advancements in the task-solving capabilities of Large Language Models (LLMs)~\cite{ouyang2022training,chen2024comm,wang2024user,huang2024survey} have motivated researchers to integrate these models into existing conversational search systems.

Most recent studies~\cite{mao2023large,ye2023enhancing}  leverage LLMs to directly generate search queries based on the context of the conversation history. Although seemingly straightforward, this technique is shown to achieve higher effectiveness in query rewriting than fine-tuning a smaller language model, such as T5~\cite{raffel2020exploring,chung2022scaling}. 
However, these performance gains are primarily achieved through the use of commercial, closed-source LLMs~\cite{chatgpt3_5,OpenAI2023GPT4TR}. This is primarily because closed-source LLMs can better perform complex reasoning tasks~\cite{gudibande2023false,kaddour2023challenges} compared to open-source models.

One of the main challenges in conversational search resides in the ambiguous nature of the conversation history. Figure~\ref{fig:intro} illustrates an example where solving co-reference relation in $u_4$ and elaborating the response in $r_3$ can help generate an adequate search query. Intuitively, performing these tasks requires basic NLP task-solving capabilities, which even 
small-scale open-source LLMs (e.g., 7B) possess~\cite{arxiv23_llama2,arxiv23_mistrial}. The key challenge is to unlock the full capabilities of open-source LLMs for conversational search. This requires carefully preparing the conversation history to enhance its quality, rather than directly using it to generate the search query.

In this paper, we propose \textbf{\fname{}}, a method that aims to enhance the quality of \textbf{\underline{c}}ontextual \textbf{\underline{h}}istory for \textbf{\underline{i}}mproving \textbf{\underline{q}}uery rewriting.
As illustrated in Figure~\ref{fig:intro}, we leverage the NLP capabilities of LLMs (e.g. solving coreference relation or expanding the context) to make the conversational history less ambiguous, consequently enhancing the relevance of the generated search query. We investigate various methods for integrating refined conversational history into existing frameworks,
including ad-hoc query rewriting, generating pseudo supervision signals for fine-tuning query rewriting models, and the fusion of both approaches.

We conduct extensive experiments using the open-source LLM, LLaMA-2-7B~\cite{arxiv23_llama2}, across five well-established conversational search benchmarks under both dense and sparse retrieval settings. 
The experimental results indicate that enhancing the conversational history using our method achieves state-of-the-art performance across most settings, often surpassing systems powered by closed-source LLMs. Our analysis reveals that although closed-source LLMs benefit from enhancing the history, the gap with open-source models is narrower when using the enhanced history with different facets compared to the original. Our contributions are summarized as follows: 
\begin{itemize}
    \item We propose a two-step method for query rewriting that relies on open-source language models: enhancing the conversation history and then generating the search query.
    \item We introduce three approaches for generating the search query on top of the enhanced conversation history: ad-hoc query rewriting (\fname{-AD}), fine-tuning a small LM for the task (\fname{-FT}), and a fusion of both approaches (\fname{-Fusion}). 
    \item Experiments conducted on five conversational search benchmarks demonstrate that \fname{}, using open-source LLMs, achieves state-of-the-art performance across most settings, often surpassing systems that rely on closed-source LLMs.  
\end{itemize}

\section{Related Work}

Different from traditional ad-hoc retrieval, which assumes users submit a stand-alone query, conversational search provides a conversational interface so that users can elaborate more complex search requirements, and interactively perform search.
The main challenge lies in accurately understanding the user's real search intent, which may be embedded within a longer, noisy, and more complex conversational context history. 
There are two well-established approaches in the literature for 
conversational search:
Conversational Dense Retrieval (CDR)~\cite{qu2020open,yu2021few,mao2024chatretriever,mo2024aligning} and Conversational Query Rewriting (CQR)~\cite{elgohary2019can}. 

The CDR systems aim to fine-tune an end-to-end conversational dense retriever that can directly model the entire conversational history to return relevant documents~\cite{kim2022saving,mo2024history}. Conversely, the CQR systems focus on formulating an adequate search query based on the conversational history. This query can then serve as the input to an existing, well-established retriever-ranker framework. We base our solution on CQR, leveraging its ability to integrate with existing ad-hoc search models, which has demonstrated significant practical value~\cite{dalton2022cast}.

Earlier approaches to CQR attempted to select useful tokens from the conversation context~\cite{kumar2020making,voskarides2020query,fang2022open} or to train generative rewriter models with conversational sessions to mimic the human-rewritten query~\cite{yu2020few,lin2020conversational,vakulenko2021question}. To optimize query rewriting, some studies have adopted reinforcement learning~\cite{wu2022conqrr,chen2022RLCQR}, or used the ranking signals with the rewriting model training~\cite{qian2022explicit,mo2023convgqr,mao2023search}. In addition, there have been endeavors to improve the conversion history quality through context denoising~\cite{lin2021contextualized,mao2022curriculum,krasakis2022zero,mo2023learning,mao2023learning}, and data augmentation~\cite{dai2022painting,mao2022convtrans,mo2024convsdg}. Unlike them, we enhance query rewriting by leveraging the NLP capabilities of open-source LLMs to reduce the ambiguity of conversational history. 
There have been multiple endeavors to integrate LLMs to solve traditional ad-hoc search sub-tasks~\cite{zhu2023large}, such as query expansion~\cite{wang2023query2doc,gao2023precise}, dense retrieval~\cite{ma2023fine,wang2024improving}, and re-ranking~\cite{sun2023chatgpt}.
About conversational search,~\citet{jin2023instructor}, \citet{jang2023itercqr}, and \citet{chen2024generalizing} attempt to improve CDR with unsupervised fine-tuning.
~\citet{mao2023large}, \citet{mo2024leverage}, and \citet{ye2023enhancing} explore how LLMs can understand users' contextualized search intents via CQR. Unlike direct rewriting the query using LLMs, we investigate various approaches for integrating refined conversational history into CQR frameworks.

\section{Methodology}
\subsection{Task Formulation}
Let $\mathcal{H}=\{u_k,r_k\}^{n}_{k=1}$ represent the user-system conversational history, where $u_k$ and $r_k$ are the user question and the system response at the $k$-th turn. 
Given a new user question $u_{n+1}$, the goal of a conversational search system is to return a set of passages $\mathcal{P}_{n+1}$ that are relevant to $\mathcal{H}$ and $u_{n+1}$, which would eventually help generate the model response $r_{n+1}$. 
To solve the main challenge of uncovering the real search intents hidden in the user's context-dependent query,
a conversational query rewriting (CQR) module has been commonly employed as an intermediate step to obtain a rewritten query $q_{n+1}$, which in turn is used as input to an off-the-shelf retriever. Recently, LLMs have become the default option for obtaining $q_{n+1}$ as follows: 
\begin{equation}
    q_{n+1} \leftarrow \mathcal{LLM}(\mathcal{I}^{CQR} \oplus \mathcal{H} \oplus u_{n+1})
\label{eq:CQR}
\end{equation}
where $\oplus$ denotes concatenation and $\mathcal{I}^{CQR}$ is a manually-engineered instruction prompt describing the CQR task. The choice of $\mathcal{LLM}(.)$ has predominantly favored closed-source commercial models, mainly \chatgpt{}, some of which require multiple iterations to get the optimal query ~\cite{mao2023large,ye2023enhancing}. In this work, we leverage the basic NLP capabilities of open-source LLMs to generate $\mathcal{H}^{\prime}$, a clearer and less noisy version of $\mathcal{H}$. This refined version can then serve as a substitute for $\mathcal{H}$ in Eq.~\ref{eq:CQR}, potentially improving the quality of $q_{n+1}^{\prime}$ using an open source LLM.

\subsection{History Enhancement}
\label{sec:History Enhancement}
In this section, we propose five approaches to tackle the ambiguity problems inherent in conversational history $\mathcal{H}$ and map each of them to a fundamental NLP task ability. Then, we explain how we design prompts for an LLM to make part or the entire history clearer. The exact prompts we used, along with illustrative examples for each case, are presented in Appendix~\ref{app:prompts}.

\subsubsection{Question Disambiguation}
\label{sec:Question Disambiguation}

Users often expect a human-to-human level of interaction with modern conversational systems. They often use acronyms, ambiguous words, or coreference substitutes when asking questions, expecting native understanding and default reasoning capabilities from these systems. For a search system, the search intent is often unclear and ambiguous. Therefore, we propose a prompt to an LLM, denoted as $ \mathcal{I}^{QD}$, that takes the conversational history $ \mathcal{H}$ and the subsequent user question $u_{n+1}$ as input, to generate $u^{\prime}_{n+1}$, a self-contained and unambiguous version of $u_{n+1}$ which can substitute it in Eq.~\ref{eq:CQR}.  

\subsubsection{Response Expansion}
\label{sec:Response Expansion}

In conversation sessions, it is common for model responses to be short and concise, especially 
for the factoid query.
While brevity is often convenient for the user to acquire needed information, it makes the response less informative for a search system, which requires abundant rewrite/expansion resources.
To handle this issue, we design a prompt $\mathcal{I}^{RE}$ which instructs an LLM to enrich the content of the last model response. The goal is to make it self-contained by leveraging the preceding conversational history. 
The enhanced history $\mathcal{H}^{\prime}$ is obtained by replacing the original response by $r^{\prime}_{n}$.

\subsubsection{Pseudo Response}
\label{sec:Pseudo Response}

Given that LLMs have been demonstrated to encapsulate human knowledge, one could employ them to speculate on potential responses directly. The intuition is that, even if the response includes some noise, it may still contain relevant terms, particularly when the LLM is prompted to produce a self-contained answer. Therefore, we design a prompt $\mathcal{I}^{PR}$ that takes the conversational history $\mathcal{H}$ and $u_{n+1}$ as inputs to generate a pseudo-response $r^{\prime}_{n+1}$. The latter can be used to expand the input of Eq.~\ref{eq:CQR} to improve the quality of the query generation.

\subsubsection{Topic Switch}
\label{sec:Topic Switch}

It is natural in a conversation that different turns may focus on different aspects. Some of them are relevant to the current turn, while others may not. This is especially the case when conversations are long.
In such cases, using the full history is highly likely to distract the CQR module, leading to poor query generation. Therefore, we design a prompt $\mathcal{I}^{TS}$ that instructs the LLM to determine whether a topic switch happens between $u_{n+1}$ and $\mathcal{H}$. 
If a switch is identified, the enhanced history would only include the last turn to maintain the transition as $\mathcal{H}^{\prime}\!=\!\{u_{n}^{\prime},r_{n}^{\prime}\}$. The other turns in $\mathcal{H}$ are deemed to be irrelevant for generating $q_{n+1}^{\prime}$ and ignored.

\subsubsection{History Summary}
\label{sec:History Summary}

As the conversation goes on, the historical context becomes longer, which contains more irrelevant/noisy parts. 
A summary of history context is expected to contain only the most useful information of the original long context and better serve for query expansion. Thus, we propose the prompt $\mathcal{I}^{HS}$, which takes a history context (original or even enhanced) and generates a summary of the conversation $\mathcal{H}^{\prime}$.

\subsection{Ad-hoc Query Rewriting}
\label{sec:Ad-hoc QR}

A straightforward method to obtain the enhanced rewritten query $q_{n+1}^{\prime}$ is to independently utilize the outputs of the five methods described previously. However, by complementing each other, the outputs of these methods can collectively contribute to a more enhanced conversational history, thereby significantly improving the retrieval performance by generating a better query.
We intuitively define multiple combinatory configurations for updating the input in Eq.~\ref{eq:CQR}, which are denoted by different symbols. 
The addition of \textbf{+QD} or \textbf{+RE} indicates that we replace $u_{n+1}$ and $r_n$ with $u_{n+1} \oplus u^{\prime}_{n+1}$ (\S~\ref{sec:Question Disambiguation}) and $r^{\prime}_n$ (\S~\ref{sec:Response Expansion}) in $\mathcal{H}^{\prime}$, respectively. \textbf{+PR} signifies that $r^{\prime}_{n+1}$ (\S~\ref{sec:Pseudo Response}) is concatenated to the input of Eq.~\ref{eq:CQR}; Lastly, \textbf{+TS} indicates that $\mathcal{H}^{\prime}$ should omit the previous turns except the last one if a topic switch is detected. Lastly,  \textbf{+HS} means that $\mathcal{H}$ is overwritten by the entire $\mathcal{H}^{\prime}$ obtained in \S~\ref{sec:History Summary}. In our default configuration, we first check for topic-switch (\textbf{TS}). If the result is affirmative, we only use the \textbf{QD+RE+PR} configuration on the top of the truncated history in \S~\ref{sec:Topic Switch}. Otherwise, we apply \textbf{QD+RE+PR} on top of the original history, followed by \textbf{+HS} which consists of obtaining the history summary on top of the enhanced history.
We report the results of models using this configuration in the remaining sections.

\subsection{Search-Oriented Fine-tuning}
\label{sec:QR Finetuning}
In existing studies, fine-tuning conversational query generators based on a small-scale language model, such as T5-base, have proven to be both effective and efficient~\cite{lin2020conversational}. 
These models consist of using human-rewritten query~\cite{wu2022conqrr} or LLM-generated query~\cite{jang2023itercqr} to serve as supervision signals and take $\mathcal{H}$ and $u_{n+1}$ as input.
However, they do not take the ranking signals into account during the training and the supervision signals might be sub-optimal~\cite{lin2021contextualized,mo2023convgqr}.

Considering that oracle search queries are typically unavailable and costly to annotate, we propose extending the existing approach to generate pseudo-supervision signals for query generation by leveraging the outputs produced in \S~\ref{sec:History Enhancement}. More precisely, we propose three modifications to Eq.~\ref{eq:CQR} to obtain a search-oriented $q_{n+1}^{\prime}$ as follow:
\begin{equation}
    Q^{\prime}_{n+1} \leftarrow \mathcal{LLM}(\hat{\mathcal{I}}^{CQR} \oplus \mathcal{H}^{\prime} \oplus u_{n+1} \oplus p_{n+1}^{*})
\label{eq:FTQR1}
\end{equation}
where we replace $\mathcal{H}$ with the enhanced history $\mathcal{H}^{\prime}$ and update the instruction $\mathcal{I}^{CQR}$ to $\hat{\mathcal{I}}^{CQR}$ to condition the query generation on the gold passage $p_{n+1}^{*}$ and prompt the LLM to generate multiple pseudo-queries in the same forward pass.
\begin{equation}
    q_{n+1}^{\prime} \leftarrow \arg \max_{q^{\prime}} \mathcal{S} (Q_{n+1}^{\prime}, p_{n+1}^{*}),\! q^{\prime} \in Q_{n+1}^{\prime}
\label{eq:FTQR2}
\end{equation}
Then, we select $q_{n+1}^{\prime}$, the one with the highest retrieval score $\mathcal{S}$ determined by an off-the-shelf retriever and relevance judgment from the set of pseudo-queries $Q^{\prime}_{n+1}$ as Eq~\ref{eq:FTQR2}, which is used as the supervision signal to fine-tune a rewriting model $\mathcal{M}(\mathcal{H}^{\prime} \oplus u_{n+1})=q_{n+1}^{\prime}$ by maximum likelihood estimation.
It is important to mention that the process described in this section is conducted offline and performed once, for the purpose of generating pseudo-labeled queries to fine-tune a search-oriented query rewriter. During inference, $\mathcal{H}$ and $u_{n+1}$ serve as inputs for the fine-tuned model to generate the query $q_{n+1}$, and no calls are made to the LLM, so that the latency is not much affected.

\section{Experimental Setup}

\subsection{Datasets and Evaluation Metrics}
To be comparable with state-of-the-art systems~\cite{mo2023convgqr,mao2023large}, we consider two standard benchmarks for conversational search: TopiOCQA~\cite{adlakha2022topiocqa}, QReCC~\cite{anantha2021open}. TopiOCQA focuses on the challenge of the topic switch under the conversational setting, while QReCC focuses on the query rewriting problem. We run experiments on the official train-test splits and report MRR, NDCG@3, and Recall@10 to evaluate the passage retrieval results as in previous works. In addition, we evaluate three CAsT datasets~\cite{dalton2020trec,dalton2021cast,dalton2022cast} which are used solely as test sets, to further validate the zero-shot or transfer learning ability of our approach, e.g., when CQR models are trained on TopiOCQA and tested on CAsTs.

\begin{table*}[!th]
\resizebox{\textwidth}{!}{
\centering
\begin{tabular}{clcccccccccccc}
\toprule
\multirow{2}{*}{{Type}} & \multirow{2}{*}{{System}} & \multicolumn{6}{c}{{System Properties}} & \multicolumn{3}{c}{TopiOCQA} & \multicolumn{3}{c}{QReCC} \\
\cmidrule(lr){3-8} \cmidrule(lr){9-11} \cmidrule(lr){12-14} & & 
DR & QR & CS & OS & FT & QF &
MRR & N@3 & R@10 & MRR & N@3 & R@10 \\
\midrule
\multirow{14}{*}{{\rotatebox{90}{\textbf{Dense (ANCE)}}}} & \texttt{ConvDR} & \cmark & \xmark & \xmark & \xmark & \xmark & \xmark & 27.2 & 26.4 & 43.5 & 38.5 & 35.7 & 58.2 \\
~ & \texttt{InstructorR}  & \cmark & \xmark & \xmark & \cmark & \xmark & \xmark & 25.3 & 23.7 & 45.1 & 43.5 & 40.5 & 66.7\\
\cmidrule(lr){2-14}
~ & \texttt{QuReTeC} & \xmark & \cmark & \xmark & \xmark & \cmark & \cmark & 11.2 & 10.5 & 20.2 & 35.0 & 32.6 & 55.0 \\
~ & \texttt{T5QR}  & \xmark & \cmark & \xmark & \xmark & \cmark & \xmark & 23.0 & 22.2 & 37.6 & 34.5 & 31.8 & 53.1 \\
~ & \texttt{CONQRR}  & \xmark & \cmark & \xmark & \xmark & \cmark & \xmark & - & - & - & 41.8 & - & 65.1 \\
~ & \texttt{ConvGQR}  & \xmark & \cmark & \xmark & \xmark & \cmark & \cmark & 25.6 & 24.3 & 41.8 & 42.0 & 39.1 & 63.5 \\
~ & \texttt{EDIRCS} & \xmark & \cmark & \xmark & \cmark & \cmark & \xmark & - & - & - & 42.1 & - & 65.6 \\
\cmidrule(lr){2-14}
~ & \texttt{IterCQR} & \xmark & \cmark & \cmark & \xmark & \cmark & \xmark & 26.3 & 25.1 & 42.6 & 42.9 & 40.2 & 65.5 \\
~ & \texttt{RETPO}$^\ddagger$  & \xmark & \cmark & \cmark & \cmark & \cmark & \cmark & 30.0 & 28.9 & 49.6 & 44.0 & 41.1 & 66.7\\
\cmidrule(lr){2-14}
~ & \texttt{LLM-Aided} & \xmark & \cmark & \cmark & \xmark & \xmark & \xmark & - & - & - & 43.9 & 41.3 & 65.6 \\
~ & \texttt{LLM4CS}  & \xmark & \cmark & \xmark & \cmark & \xmark & \cmark & 27.7 & 26.7 & 43.3 & 44.8 & 42.1 & 66.4 \\
\cmidrule(lr){2-14}
~ & \texttt{\fname{-FT}} & \xmark & \cmark & \xmark & \cmark & \cmark & \xmark & 30.0$^\dagger$ & 28.9$^\dagger$ & 51.0$^\dagger$ & 36.9 & 34.0 & 57.6 \\
~ & \texttt{\fname{-AD}}  & \xmark & \cmark & \xmark & \cmark & \xmark & \xmark & \underline{33.2}$^\dagger$ & \underline{32.2}$^\dagger$ & \underline{53.0}$^\dagger$ &  \underline{47.0}$^\dagger$ & \textbf{44.6}$^\dagger$ & \textbf{70.8}$^\dagger$ \\
~ & \texttt{\fname{-Fusion}} & \xmark & \cmark & \xmark & \cmark & \cmark & \cmark & \textbf{38.0}$^\dagger$ & \textbf{37.0}$^\dagger$ & \textbf{61.6}$^\dagger$ & \textbf{47.2}$^\dagger$ & \underline{44.2}$^\dagger$ & \underline{70.7}$^\dagger$ \\
\midrule
\multirow{11}{*}{{\rotatebox{90}{\textbf{Sparse (BM25)}}}} & \texttt{QuReTeC} & \xmark & \cmark & \xmark & \xmark & \cmark & \cmark & 8.5 & 7.3 & 16.0 & 34.0 & 30.5 & 55.5 \\
~ & \texttt{T5QR} & \xmark & \cmark & \xmark & \xmark & \cmark & \xmark & 11.3 & 9.8 & 22.1 & 33.4 & 30.2 & 53.8 \\
~ & \texttt{CONQRR} & \xmark & \cmark & \xmark & \xmark & \cmark & \xmark &  - & - & - & 38.3 & - & 60.1 \\
~ & \texttt{ConvGQR} & \xmark & \cmark & \xmark & \xmark & \cmark & \cmark &  12.4 & 10.7 & 23.8 & 45.6 & 44.1 & 41.0 \\
~ & \texttt{EDIRCS} & \xmark & \cmark & \xmark & \cmark & \cmark & \xmark &  - & - & - & 41.2 & - & 62.7  \\
\cmidrule(lr){2-14}
~ & \texttt{IterCQR} & \xmark & \cmark & \cmark & \xmark & \cmark & \xmark &  16.5 & 14.9 & 29.3 & 46.7 & 44.1 & 64.4 \\
\cmidrule(lr){2-14}
~ & \texttt{LLM-Aided} & \xmark & \cmark & \cmark & \xmark & \xmark & \xmark &  - & - & - & 48.9 & 46.3 & 66.4 \\
~ & \texttt{LLM4CS} & \xmark & \cmark & \xmark & \cmark & \xmark & \cmark & 18.9 & 17.7 & 33.7 & 47.8 & 45.0 & 69.1 \\
\cmidrule(lr){2-14}
~ & \texttt{\fname{-FT}} & \xmark & \cmark & \xmark & \cmark & \cmark & \xmark &  17.0$^\dagger$ & 15.4$^\dagger$ & 32.3$^\dagger$ & 37.8 & 35.0 & 57.1 \\
~ & \texttt{\fname{-AD}} & \xmark & \cmark & \xmark & \cmark & \xmark & \xmark &  \underline{22.5}$^\dagger$ & \underline{20.5}$^\dagger$ & \underline{40.4}$^\dagger$ & \underline{53.1}$^\dagger$ & \underline{50.7}$^\dagger$ & \underline{77.2}$^\dagger$ \\
~ & \texttt{\fname{-Fusion}} & \xmark & \cmark & \xmark & \cmark & \cmark & \cmark & \textbf{25.6}$^\dagger$ & \textbf{23.5}$^\dagger$ & \textbf{44.7}$^\dagger$ & \textbf{54.3}$^\dagger$ & \textbf{51.9}$^\dagger$ & \textbf{78.5}$^\dagger$ \\
\bottomrule
\end{tabular}
}
\caption{Performance of dense and sparse retrieval on TopiOCQA and QReCC with different systems. 
We list the attributes of the reported baseline systems, which include: \underline{\bf DR} based on conversational dense retrieval, \underline{\bf QR} perform query rewriting, \underline{\bf CS} leverage close-source LLMs (e.g., ChatGPT or GPT-4), \underline{\bf OS} leverage open-source LLMs (mainly LLaMA-2-7B), \underline{\bf FT} fine-tune a small LM (mainly T5-base) for QR, and \underline{\bf QF} fuse multiple queries for retrieval. \texttt{RETPO}$^\ddagger$ involves high-cost supervised fine-tuning an LLM for QR. $\dagger$ denotes significant improvements with t-test at $p<0.05$ over all compared baselines (except \texttt{CONQRR}, \texttt{RETPO}, and \texttt{IterCQR}). \textbf{Bold} and \underline{underline} indicate the best and the second-best results within the categories of dense and sparse retrieval.}
\label{tab:main_res}
\end{table*}

\subsection{Baselines}

We define three main configurations for the approaches that use our enhanced generated queries:

\begin{itemize}
    \item \textbf{\fname{-AD}} directly use the queries generated by the ad-hoc method described in \S~\ref{sec:Ad-hoc QR} as input for an off-the-shelf retriever.

    \item \textbf{\fname{-FT}} search queries generated by a small LM (e.g., T5) fine-tuned for the CQR task following the approach described in \S~\ref{sec:QR Finetuning}. 

    \item \textbf{\fname{-Fusion}} We fuse the rank list retrieved
    by \textit{\fname{-AD}} and \textit{\fname{-FT}} using the \textit{result-level} fusion technique~\cite{lin2021contextualized}.\footnote{It consists of aggregating multiple ranked lists retrieved by each query into a single list to produce $\mathcal{P}_{n+1}$.}
\end{itemize}

We compared our methods with a variety of systems that can mainly be classified into three categories. More precisely, we first compare against traditional systems that fine-tune small-scale CQR models (e.g., T5-base) including: 
\texttt{QuReTeC}~\cite{voskarides2020query}
\texttt{T5QR}~\cite{lin2020conversational},  \texttt{CONQRR}~\cite{wu2022conqrr}, \texttt{ConvGQR}~\cite{mo2023convgqr}, \texttt{EDIRCS}~\cite{mao2023search}.
Then, we compare with the systems that fine-tune an LLM-based retriever, e.g., create the query and document representation by the ending token from a decoder-only model, including RepLLaMA~\cite{ma2023fine}, \texttt{E5-Mistral}~\cite{wang2024improving}, and \texttt{LLM-Embedder}~\cite{zhang2023retrieve}, or fine-tune an LLM-based CQR model as \texttt{RETPO}~\cite{yoon2024ask} and \texttt{IterCQR}~\cite{jang2023itercqr}.
Besides, we include the systems that directly obtain the rewritten query by prompting LLMs such as \texttt{LLM-Aided IQR}~\cite{ye2023enhancing}, \texttt{HyDE}~\cite{gao2023precise}, \texttt{Query2doc}~\cite{wang2023query2doc} and \texttt{LLM4CS}~\cite{mao2023large}.
Although not directly comparable, we report results of systems that fine-tune an ad-hoc search retriever for conversational scenarios, including the one without LLMs \texttt{ConvDR}~\cite{yu2021few} and with LLMs \texttt{InstructorR}~\cite{jin2023instructor}. 
A detailed description of each aforementioned baseline is presented in Appendix~\ref{app:Baselines}.

\subsection{Implementation Details}
We conduct experiments with the instruct-tuning variants\footnote{Concretely, the version of LLMs we used are \textit{meta-llama/Llama-2-7b-chat-hf} and \textit{mistralai/Mistral-7B-Instruct-v0.2} on \url{https://huggingface.co/}, respectively.} of both LLaMA-2-7B~\cite{arxiv23_llama2} and Mistral-2-7B~\cite{arxiv23_mistrial} as $\mathcal{LLM(.)}$ in Eq.~\ref{eq:CQR} and Eq.~\ref{eq:FTQR1}. We experiment with both BM25~\cite{robertson2009probabilistic} sparse retriever and ANCE dense retriever~\cite{xiong2020approximate}.  
In addition, we use FlanT5-base\footnote{We report our main results using FlanT5-base to ensure the results are comparable with previous studies.}~\cite{chung2022scaling} and large models as the backbone when fine-tuning a CQR model on TopiOCQA and QReCC. The fine-tuning process consists of 10 epochs with a learning rate of 1e-5 and a batch size of 8 for both datasets. 
More implementation details can be found in Appendix~\ref{app:Implementation Details}.

\begin{table*}[!th]
\centering
\resizebox{\textwidth}{!}{
\begin{tabular}{lcccccccccc}
\toprule
\multicolumn{1}{l}{\multirow{2}{*}{System}} & \multirow{2}{*}{$\mathcal{LLM}$} & \multicolumn{3}{c}{CAsT-19} & \multicolumn{3}{c}{CAsT-20} & \multicolumn{3}{c}{CAsT-21} \\ 
\cmidrule(lr){3-5} \cmidrule(lr){6-8} \cmidrule(lr){9-11} 
& & MRR & N@3  & R@10  & MRR & N@3 & R@10 & MRR     & N@3 & R@10  \\ 
\midrule
\texttt{RepLLaMA} & LLaMA-2-7B & 62.4 & 31.6 & 10.6 & 26.8 & 18.3 & 10.4 & 47.4 & 32.7 & 19.6 \\
\texttt{E5-Mistral} & Mistral-7B & 62.2 & 31.3 & 9.5 & 22.0 & 15.4 & 8.4 & 48.2 & 32.5 & 20.5\\
\texttt{LLM-Embedder} & LLaMA-2-7B & 63.3 & 36.6 & 11.4 & 25.2 & 15.4 & 8.7 & 46.8 & 31.2 & 17.3\\
\texttt{HyDE} & ChatGPT-3.5 & 55.6 & 39.2 & 10.0 & 44.8 & 29.3 & 16.9 & - & - & -\\
\texttt{Query2doc} & ChatGPT-3.5 & 58.8 & 42.4 & 11.6 & 48.6 & 32.5 & 17.3 & - & - & -\\
\texttt{InstructorR} & ChatGPT-3.5 & 61.2 & 46.6 & 10.4 & 43.7 & 29.6 & 8.3 & 46.7 & 32.5 & 18.4 \\
\multirow{3}{*}{\texttt{LLM4CS}} & LLaMA-2-7B & 68.4 & 45.9 & 11.2 & 52.3 & 37.2 & \underline{17.9} & 57.0 & 41.5 & 20.2\\
& Mistral-2-7B & 67.6 & 44.5 & 10.9 & 48.3 & 33.5 & 17.0 & 53.0 & 35.3 & 19.6\\
& ChatGPT-3.5 & 70.4 & 46.8 & 11.7 & \textbf{58.6} & \textbf{41.5} & \textbf{19.3} & \bf 66.1 & \bf 46.9 & \underline{24.4}\\
\midrule
\texttt{\fname{-FT}} & LLaMA-2-7B & 68.5 & 45.1 & \underline{11.9}$^\dagger$ & 46.3 & 31.6 & 15.9 & 53.9 & 36.0 & 20.4\\
\texttt{\fname{-AD}} & LLaMA-2-7B & \underline{70.8}$^\dagger$ & 47.6$^\dagger$ & \underline{11.9}$^\dagger$ & 51.0 & 34.4 & \underline{17.9} & 57.7 & 42.0 & 22.6\\
\texttt{\fname{-Fusion}} &  LLaMA-2-7B & \bf 73.3$^\dagger$ & \bf 50.5$^\dagger$ & \bf 12.9$^\dagger$ & \underline{54.0} & \underline{38.0} & \bf 19.3 & \underline{62.9} & \underline{46.5} & \bf 25.2$^\dagger$ \\
        \bottomrule
     \end{tabular}}
     \caption{Zero-shot retrieval performances of the systems involved with LLMs under the dense retrieval (ANCE).
     $\dagger$ denotes significant improvements with t-test at $p<0.05$ over all compared baselines. \textbf{Bold} and \underline{underline} indicate the best and the second-best results, respectively.}
     \label{tab:cast_main}
\end{table*}

\section{Results and Analysis}

\subsection{Main Results}
\label{sec:Main Results}

\autoref{tab:main_res} shows both the dense and sparse retrieval performances of systems with diverse properties on the TopiOCQA and QReCC. We report the results of our systems using LLaMA-2-7B as the backbone LLM to make the results comparable with previous work. 
First, we observe that using our enhanced conversation history significantly improves performance over vanilla baselines that use the original history, for both ad-hoc QR (\texttt{LLM4CS}) and fine-tuning a small QR model (\texttt{T5QR}). For dense retrieval, \texttt{\fname{-AD}} outperforms \texttt{LLM4CS} by 5.5\% and 2.2\% MRR on TopiOCQA and QReCC respectively, while \texttt{\fname{-FT}} reports a gain of 7.0\% and 1.9\% over \texttt{T5QR} on the same datasets. 
Similar gains are also observed using the sparse retriever, indicating the strong effectiveness of our methods.

Second, we notice that vanilla QR systems on top of an enhanced history can outperform systems that utilize additional training techniques and sophisticated modules. For instance, \texttt{\fname{-AD}} outperforms both \texttt{ConvDR} and \texttt{IntructorR}, which need relevance judgments to fine-tune a conversational dense retriever on the raw input; \texttt{ReTPO}, which fine-tunes an LLM for QR and in addition leverages GPT-4 for data augmentation. 
While \texttt{\fname{-FT}} outperforms its direct competitors, primarily \texttt{ConvGQR} and \texttt{IterCQR} that refine the supervision signals on TopiOCQA, it underperforms on QReCC mainly because previous fine-tuned QR models rely on QReCC's human-rewritten queries. The contrasting observations between the two datasets suggest that enhancing the history is crucial for performance when no QR-supervised annotations exist.

Third, we observe that by systematically fusing the outputs of our approaches,
\texttt{\fname{-Fusion}} outperforms the models that use each component separately, 
achieving the best performance across most settings. The gains are more significant on the topic-mixed and more challenging TopiOCQA compared to QReCC, with 4.8\% and 0.2\% MRR score improvements on each dataset. Interestingly, this occurs even though \texttt{\fname{-FT}} significantly underperforms compared to \texttt{\fname{-AD}}, suggesting that \texttt{\fname{-FT}} still generates query content that is complementary and not captured by \texttt{\fname{-AD}}.

\subsection{Zero-shot Results}
\label{sec:Zero-shot Results}

We compare the dense retrieval performances of different systems that leverage LLMs under a zero-shot manner on three CAsT datasets in \autoref{tab:cast_main}. 
We observe a consistent pattern as previous results in \autoref{tab:main_res} when comparing the performances within our approaches. More precisely, although \texttt{\fname{-FT}} performs slightly worse compared to \texttt{\fname{-AD}}, fusing their outputs systematically leads to better performances across all three datasets.
Besides, we can see that \texttt{\fname{-AD}} outperforms most systems either utilizing open-source or close-source LLMs and yields results competitive with the state-of-the-art system \texttt{LLM4CS}, which requires multiple calling for each query turn. Specifically, \texttt{\fname{-AD}} surpasses \texttt{LLM4CS} with LLaMA-2-7B, on CAsT-19 and CAsT-21. 
In addition, our top-performing approach, \texttt{\fname{-Fusion}}, outperforms all compared systems, except the \texttt{LLM4CS} with close-source ChatGPT-3.5 on CAsT-20 and CAsT-21, indicating the superior effectiveness of our approaches.
We also find that ad-hoc fine-tuned LLM-based retrievers (\texttt{RepLLaMA}, \texttt{E5-Mistral}, and \texttt{LLM-Embedder}) underperform the systems with LLM-based query generation (\texttt{HyDE} and \texttt{Query2doc}) and the \texttt{InstructorR} with conversational fine-tuning adaption. 
These systems also underperform \texttt{\fname{-FT}}, which only fine-tunes a small LM on TopiOCQA with enhanced supervision signals. The observation indicates the importance of improving the generalization capabilities of the models to handle complex and diverse conversational scenarios.

\subsection{Open vs. Close Source LLMs}
In addition to conducting experiments with open-source LLMs, we also deploy the closed-source LLM ChatGPT-3.5 to isolate the effects of history enhancement. \autoref{tab:open_close_llm} shows the dense retrieval performances on three CAsT test sets when query rewriting (QR) is performed by \texttt{\fname{-AD}} approach on both the original history and the enhanced one. We observe that ChatGPT-3.5 benefits from QR on enhanced history, with NDCG@3 score improvements of 5.4\%, 4.3\%, and 0.9\% across CAsT-19, CAsT-20, and CAsT-21, respectively. Such results indicate that despite the superior reasoning abilities of closed-source LLM, enhancing the conversational history is deemed important for handling complex queries within conversational scenarios. 

\begin{table}[!th]
\centering
\resizebox{\columnwidth}{!}{
\begin{tabular}{lcccccc}
\toprule
\multirow{2}{*}{$\mathcal{LLM}$} & \multicolumn{2}{c}{CAsT-19} & \multicolumn{2}{c}{CAsT-20} & \multicolumn{2}{c}{CAsT-21}\\
\cmidrule(lr){2-3}\cmidrule(lr){4-5}\cmidrule(lr){6-7}
~ & {MRR} & {N@3} & {MRR} & {N@3} & {MRR} & {N@3} \\
\midrule
\multicolumn{7}{c}{Original History} \\
\midrule
LLaMA & 67.4 & \bf 42.5 & 40.9 & 27.9 & 52.7 & 37.2 \\
Mistral & 67.9 & 42.0 & 44.2 & 30.4 & 59.5 & 41.6 \\
ChatGPT & \bf 69.3 & 40.8 & \bf 53.0 & \bf 36.2 & \bf 60.3 & \bf 41.9 \\
\midrule
\multicolumn{7}{c}{Our Enhanced History}  \\
\midrule
LLaMA & 70.8 & \bf 47.6 & 51.0 & 34.4 & 57.7 & 42.0\\
Mistral & 71.4 & 47.2 & 49.2 & 34.4 & \bf 67.0 & \bf 47.2\\
ChatGPT & \bf 71.7 & 46.4 & \bf 55.7 & \bf 40.5 & 62.2 & 42.8\\
\bottomrule
\end{tabular}
}
\caption{Dense retrieval results for systems using various LLMs as backbones, where QR is performed either directly on top of the original conversation history or on our enhanced history using the \texttt{\fname{-AD}} method.}
\label{tab:open_close_llm}
\vspace{-2ex}
\end{table}

Also, we find that conducting QR on enhanced conversational history helps to narrow the performance gap between open-source and closed-source LLMs. For instance, the gap of MRR score between LLaMA and ChatGPT-3.5 on the original history is 1.9\%, 12.1\%, and 7.6\% across three CAsT test sets, respectively. 
In contrast, when utilizing enhanced history, the gaps reduced significantly to 0.9\%, 4.7\%, and 4.5\%, indicating that our designed approach can adequately leverage the capacity of open-source LLMs for conversational search and be competitive with close-source ones.

\subsection{Search-Oriented Fine-tuning Ablation}
\label{sec:Finetuning Ablation}

We analyze the potential choices for generating search queries in Eq.~\ref{eq:FTQR1} and Eq.~\ref{eq:FTQR2} as supervision signals for \texttt{\fname{-FT}} models. \autoref{tab:finetune_abl} presents the dense retrieval performances via the queries generated by \texttt{\fname{-FT}} models, which are fine-tuned using manually rewritten queries or the variants of the approach outlined in \S \ref{sec:QR Finetuning}.
The ablations are based on the results of either without using enhanced history, without generating multiple queries, or not conditioning on the gold passage. We observe that using the queries generated by LLMs as supervision signals outperform the one using manual annotation, which is consistent with previous studies that have identified human-written queries as sub-optimal \cite{wu2022conqrr,mo2023convgqr}.

\begin{table}[!th]
\centering
\scalebox{0.87}{
\begin{tabular}{lcccc}
\toprule
\multirow{2}{*}{Signal} &\multicolumn{2}{c}{TopiOCQA} & \multicolumn{2}{c}{QReCC} \\
\cmidrule(lr){2-3}\cmidrule(lr){4-5}
~ & {MRR} & {N@3} & {MRR} & {N@3} \\
\midrule
Manual & - & - & 34.7 & 31.9\\
\midrule
\fname{-FT} & 30.0 & 28.9 & 36.9 & 34.0 \\
\hspace{2mm} w.o. $\mathcal{H}^{\prime}$ & 27.6 & 26.7 & 35.4 & 33.8 \\
\hspace{2mm} w.o. $Q_{n+1}^{\prime}$ & 24.2 & 23.4 & 33.4 & 31.7 \\
\hspace{2mm} w.o. $p_{n+1}^{*}$ & 18.2 & 17.2 & 26.8 & 23.9\\
\bottomrule
\end{tabular}
}
\caption{Dense retrieval performances of fine-tuned QR models, which utilize different supervised signals. These include human-written queries and variants from \texttt{\fname{-FT}}, either without enhanced history, without multiple queries, or not conditioned on the gold passage.}
\label{tab:finetune_abl}
\vspace{-2.5ex}
\end{table}
Additionally, we observe that all of our proposed adaptive modifications significantly enhance the final performance of the system, especially applying them all to generate the pseudo supervision signals. 
Such results indicate that improving the quality of supervision signals is crucial for QR model fine-tuning and justifying the effectiveness of our approaches for search-oriented fine-tuning.

\subsection{History Enhancement Ablation}
\label{sec:History Enhancement Ablation} 

We study the contribution of each of our proposed prompts in \S~\ref{sec:History Enhancement} for query enhancement by conducting ablation studies on $\mathcal{H}^{\prime}$ when one of the prompts is not used. \autoref{tab:history_abl} presents the ablation performances of dense and sparse retrieval on TopiOCQA and QReCC. 
The History Summary (\textbf{HS}) is ablated alongside (\textbf{TS}), as \textbf{HS} is not activated if a new topic is detected.

\begin{table}[!th]
\centering
\resizebox{\columnwidth}{!}{
\begin{tabular}{clcccc}
\toprule
\multirow{2}{*}{Type} &\multirow{2}{*}{Ablation} &\multicolumn{2}{c}{TopiOCQA} & \multicolumn{2}{c}{QReCC} \\
\cmidrule(lr){3-4}\cmidrule(lr){5-6}
~ & ~ & {MRR} & {N@3} & {MRR} & {N@3} \\
\midrule
\multirow{5}{*}{{\begin{sideways}\specialcell{\bf Dense\\(ANCE)}\end{sideways}}} & \fname{-AD} & 33.2 & 32.2 & 47.0 & 44.6 \\
& \hspace{2mm} w.o. QD & 32.5 & 31.4 & 44.6 & 41.9\\
& \hspace{2mm} w.o. RE & 28.3 & 27.0 & 46.6 & 44.0\\
& \hspace{2mm} w.o. PR & 26.4 & 25.2 & 43.5 & 40.8\\
& \hspace{2mm} w.o. TS & 20.0 & 18.7 & 46.9 & 44.4\\
\midrule
\multirow{5}{*}{{\begin{sideways}\specialcell{\bf Sparse\\(BM25)}\end{sideways}}} & \fname{-AD} & 22.5 & 20.5 & 53.1 & 50.7 \\
& \hspace{2mm} w.o. QD & 22.1 & 20.1 & 47.3 & 44.6 \\
& \hspace{2mm} w.o. RE & 19.0 & 16.8 & 50.1 & 47.3\\
& \hspace{2mm} w.o. TS & 17.8 & 16.5 & 51.7 & 48.8\\
& \hspace{2mm} w.o. PR & 16.9 & 15.3 & 46.9 & 44.5\\
\bottomrule
\end{tabular}
}
\caption{Dense and sparse retrieval results of ablating \texttt{\fname{-AD}} by not using one history enhancement prompt at each line on TopiOCQA and QReCC datasets.}
\label{tab:history_abl}
\vspace{-2ex}
\end{table}

We observe that all our proposed enhancements to the history context contribute positively to the performance of the \texttt{\fname{-AD}} method, although some enhancements are more effective than others. On one hand, detecting topic switching is particularly crucial on TopiOCQA, leading to performance improvements of 13.2\% and 5.2\% MRR scores in dense and sparse retrieval, respectively. This is mainly due to the multi-topic focus design of the dataset within the same conversation. On the other hand, we notice that while question disambiguation \textbf{(QD)} improves performance, it is less critical compared to predicting a pseudo response \textbf{(PR)} or enhancing the quality of the last system response \textbf{(RE)}. In addition, we notice that all our proposed enhancements contribute similarly to the generated search queries across both dense and sparse retrieval settings.

\subsection{Case Analysis}
\label{sec:Case Analysis}

We manually analyze the content of the enhanced history to better understand the mechanisms and limitations of our approach. This analysis shows the complementary roles each enhancement prompt plays in improving the quality of the original history. \textbf{QD} and \textbf{RE} primarily assist in resolving coreferences and clarifying acronyms to full names, \textbf{TS} helps remove irrelevant content, \textbf{PR} speculates on relevant terms that may occur in the response, and \textbf{HS} not only converts the conversation into plain text but also ensures that key terms from the conversation are preserved. 
While prompts such as \textbf{PR} and \textbf{RE} generation generally aid in retrieval, they may also introduce noisy terms due to the wrong fact generated by LLMs that hurt the ranking results.
Finally, we also notice that the queries generated by \texttt{\fname{-AD}} and \texttt{\fname{-FT}} are of different styles. The first focuses on expanding more relevant terms to increase the matching scores, while the latter queries are more concise with higher efficiency for retrieval. Nevertheless, aggregating the output rank lists from both approaches helps refine the final results by ranking the relevant passages higher. The concrete examples of these case analyses are presented in Appendix~\ref{app:Case Analysis}.

\section{Conclusion}
In this paper, we propose \fname{}, an approach that leverages the basic NLP capabilities of LLMs to enhance the quality of contextual history 
for improving the query rewriting performance in terms of conversational search. Despite its simplicity, our approach achieves superior performance across various datasets and settings, using open-source LLMs compared to closed-source alternatives. This study shows that instead of simply ask an LLM to generate a search query, it is critical to design strategies to generate different facets of enhancement in view of finding the target information.

\section*{Limitations}
Potential limitations of this work include not experimenting with larger open-source LLMs, such as the 56B Mixtral~\cite{mixtral} or 70B LLaM
a, as well as other recent models like Gemma~\cite{team2024gemma}. Additionally, the study did not incorporate more closed-source models such as GPT-4~\cite{OpenAI2023GPT4TR} or Claude~\cite{claude} to further study the impact of history enhancement. This is mainly due to limitations in computation (open source) and financial (close source) resources.  Despite the straightforward and significant gains, some design choices could be further analyzed to potentially boost the performance even more. For instance, adding a backoff filtering strategy could detect when the LLM is producing noisy outputs, or exploring approaches that interpolate between the use of human and pseudo-queries when its higher quality as training signals for \texttt{\fname{-FT}}. Besides, we have considered 5 directions of enhancement in this paper. More strategies can be incorporated so that other useful enhancements can be integrated.



\bibliography{anthology,custom}
\bibliographystyle{acl_natbib}

\clearpage
\appendix

\section{History Enhancement and Query Rewriting Prompts}
\label{app:prompts}
In this section, we list the prompts that we have carefully designed to enhance different parts of the conversation history, as well as the prompt used for query rewriting and pseudo supervision signals for search-originated fine-tuning. For each prompt, we designed the instruction part through trial and error iterations until we confirmed that both models (LLaMA-2-7B and Mistral-v0.2-7B) could follow the instructions and generate outputs in the required format. We observed no benefits from using in-context examples for any model, as the outputs remained mostly stable, with minor to no changes in the model responses even after adding these examples.

\subsection{Question Disambiguation}

{\small \texttt{You are given a set of question-answers pairs and a new question that is ambiguous. Your goal is to rewrite the question so it becomes clear. Write the new question without any introduction.}
}
\normalsize

\subsection{Response Expansion}
{\small \texttt{Given a series of question-and-answer pairs, along with a new question, your task is to give a one-sentence response to the new question.}
}
\normalsize

\subsection{Pseudo Response}

{\small \texttt{You are given a question-and-answer pair, where the answer is not clear. Your goal is to write a long version of the answer based on its given context. The generated answer should be one sentence only and less than 20 words.}
}
\normalsize

\subsection{Topic Switch}

{\small \texttt{Given a series of question-and-answer pairs, along with a new question, your task is to determine whether the new question continues the discussion on an existing topic or introduces a new topic. Please respond with either "new\_topic" or "old\_topic" as appropriate.}
}
\normalsize

\subsection{History Summary}

{\small \texttt{You are given a context in the form of question-answer pairs. Your goal is to write a paragraph that summarizes the information in the context. The summary should be short with one sentence for each question answer pair.}
}
\normalsize

\subsection{Query Rewriting}

{\small \texttt{Given a series of question-and-answer pairs as context, along with a new question, your task is to convert the new question into a search engine query that can be used to retrieve relevant documents. The output should be placed in a JSON dictionary as follows: \{"query": ""\}}
}
\normalsize

\subsection{Pseudo Supervision Signals}

{\small \texttt{You are given a relevant passage, a series of question-and-answer pairs as context along with a new question, your task is to generate a set of search queries based on the relevancy between the new question and the relevant passage and also rely on the given context. The output format should be in a list with indexes e.g., 1. 2. 3.
}}
\normalsize

\section{Experimental Setup}
\label{app:Experimental Setup}

\subsection{Datasets Details}
\label{appendix: datasets}

\begin{table}[!th]
\centering
\small
\setlength{\tabcolsep}{4pt}{
\begin{tabular}{llrrr}
\toprule
Dataset & Split & \#Conv. & \#Turns(Qry.) & \#Collection \\ \midrule
\multirow{2}{*}{TopiOCQA} & Train & 3,509 & 45,450 & \multirow{2}{*}{25M} \\
 & Test  & 205 & 2,514 & \\
\midrule
\multirow{2}{*}{QReCC} & Train & 10,823 & 29,596 & \multirow{2}{*}{54M} \\
 & Test  & 2,775 & 8,124 & \\
\midrule
CAsT-19 & Test & 50 & 479 & \multirow{2}{*}{38M}\\
CAsT-20 & Test & 25 & 208 & \\
\midrule
CAsT-21 & Test & 26 & 239 & 40M \\
\bottomrule
\end{tabular}}
\caption{Statistics of conversational search datasets.}
\label{table: datasets}
\end{table}
The statistics of each dataset are presented in Table~\ref{table: datasets}. We discard the samples without gold passages.
The manually rewritten query for each turn is provided in all datasets except TopiOCQA.
The relevance judgments in the CAsT datasets are made by experts with multi-level annotations. The relevance judgment thresholds are set at 1, 2, and 2 for CAsT-19, CAsT-20, and CAsT-21, respectively.

\subsection{Implementation Details}
\label{app:Implementation Details}

We implement the retrieval evaluation metrics from the \texttt{pytrec\_eval} tool~\cite{sigir18_pytrec_eval}. We leverage the Pyserini~\cite{lin2021pyserini} and Faiss~\cite{johnson2019billion} libraries for implementing the BM25 and ANCE retrievers, respectively. Following previous works~\cite{lin2021contextualized,mo2023convgqr}, we set BM25 parameters as follows: $k_1=0.9, b=0.4$ on TopiOCQA and $k_1=0.82, b=0.68$ on the QReCC. The lengths of the query, concatenated input, and passage are truncated to 32, 512, and 384 tokens, respectively.

In all experiments, we use sampling and set the temperature to 0.7 when generating with LLMs. For the search-oriented fine-tuning, we use the NDCG@3 score as the standard metric to select the generated query as the pseudo supervision signals, while we set the maximum length for the generated query is set to 32, which is the same as~\cite{lin2020conversational,wu2022conqrr,mo2023convgqr}.
For the rank-list fusion, we set the balance factor $\alpha$ in \citet{lin2021contextualized} as 1, which indicates the same importance of different retrieved results.

\begin{table*}[!t]
\centering
\begin{tabular}{llcccccc}
\toprule
\multirow{2}{*}{Type} & \multirow{2}{*}{Systems} & \multicolumn{3}{c}{TopiOCQA} & \multicolumn{3}{c}{QReCC} \\
\cmidrule(lr){3-5}\cmidrule(lr){6-8}
~ & ~ & {MRR} & {NDCG@3} & Recall@10 & {MRR} & {NDCG@3} & Recall@10 \\
\midrule
\multirow{3}{*}{{\begin{sideways}\specialcell{\bf Dense}\end{sideways}}} & \texttt{\fname{-FT}} & 26.2 & 25.4 & 45.3 & 31.1 & 28.5 & 50.6 \\
~ & \texttt{\fname{-AD}} & 28.9$^\dagger$ & 28.3$^\dagger$ & 46.8$^\dagger$ & 46.7$^\dagger$ & \textbf{44.2}$^\dagger$ & \textbf{70.7}$^\dagger$ \\
~ & \texttt{\fname{-Fusion}} & \textbf{36.3}$^\dagger$ & \textbf{35.0}$^\dagger$ & \textbf{59.6}$^\dagger$ & \textbf{47.1}$^\dagger$ & 44.1$^\dagger$ & 70.3$^\dagger$ \\
\midrule
\multirow{3}{*}{{\begin{sideways}\specialcell{\bf Sparse}\end{sideways}}} & \texttt{\fname{-FT}} & 15.2 & 14.3 & 30.5 & 32.9 & 30.0 & 51.4 \\
~ & \texttt{\fname{-AD}} & 19.2$^\dagger$ & 17.3$^\dagger$ & 35.6$^\dagger$ & 51.7$^\dagger$ & 48.8$^\dagger$ & 76.2$^\dagger$ \\
~ & \texttt{\fname{-Fusion}} & \textbf{21.4}$^\dagger$ & \textbf{19.2}$^\dagger$ & \textbf{39.4}$^\dagger$ & \textbf{51.9}$^\dagger$ & \textbf{49.0}$^\dagger$ & \textbf{76.3}$^\dagger$ \\
\bottomrule
\end{tabular}
\caption{Performance of dense and sparse retrieval on TopiOCQA and QReCC datasets based on Mistral-2-7B model. The system properties and the settings are inherited from the \autoref{tab:main_res}.}
\label{tab:main_mistral}
\end{table*}

\begin{table*}[!t]
\resizebox{\textwidth}{!}{
\centering
\begin{tabular}{lccccccccc}
\toprule
\multirow{2}{*}{Systems} & \multicolumn{3}{c}{CAsT-19} & \multicolumn{3}{c}{CAsT-20} & \multicolumn{3}{c}{CAsT-21}\\
\cmidrule(lr){2-4}\cmidrule(lr){5-7}\cmidrule(lr){8-10}
~ & {MRR} & {NDCG@3} & R@10 & {MRR} & {NDCG@3} & R@10 & {MRR} & {NDCG@3} & R@10 \\
\midrule
\texttt{\fname{-FT}} & 58.3 & 35.4 & 9.0 & 37.1 & 24.7 & 12.0 & 44.4 & 29.1 & 16.5\\
\texttt{\fname{-AD}} & 71.4$^\dagger$ & 47.2$^\dagger$ & \textbf{12.2}$^\dagger$ & 49.2 & 34.4 & 17.5 & \textbf{67.0}$^\dagger$ & 47.2$^\dagger$ & 25.5$^\dagger$\\
\texttt{\fname{-Fusion}} & \textbf{71.5}$^\dagger$ & \textbf{47.7}$^\dagger$ & 11.9$^\dagger$ & \textbf{51.1} & \textbf{35.5} & \textbf{18.4} & 66.1 & \textbf{48.9}$^\dagger$ & \textbf{27.5}$^\dagger$\\
\bottomrule
\end{tabular}
}
\caption{Performance of dense retrieval on three CAsT datasets based on Mistral-2-7B model. The system properties and the settings are inherited from the \autoref{tab:cast_main}.}
\label{tab:cast_mistral}
\end{table*}

\subsection{Baselines Details}
\label{app:Baselines}
We provide a more detailed introduction to the following baselines used for comparison:

\textbf{QuReTeC}~\cite{voskarides2020query}: A traditional sequence tagger query rewriting approach fine-tuned with weakly supervision signals to determine whether a term in a historical context should be expanded to the current query.

\textbf{T5QR}~\cite{lin2020conversational}: A query rewriting approach fine-tuned with manual annotations provided in QReCC as the supervised signals via the T5-base model.

\textbf{CONQRR}~\cite{wu2022conqrr}: A query rewriting approach fine-tuned with manual annotations provided in QReCC and the ranking signals using reinforcement learning via the T5-base model.

\textbf{ConvGQR}~\cite{mo2023convgqr}: A unified framework that integrates query rewriting and query expansion mechanisms by two T5-base models and fine-tuned them with manual annotations and ground-truth response, respectively.

\textbf{EDIRCS}~\cite{mao2023search}: A query rewriting approach based on text editing technique with ranking signals fine-tuned on the T5-base model.

\textbf{IterCQR}~\cite{jang2023itercqr}: An iterative query rewriting method using the initial query generated by the ChatGPT-3.5 and refining a query turn multiple times according to the ranking signals feedback during the training stage.

\textbf{LLM-Aided}~\cite{ye2023enhancing}: An informative conversational query rewriting by directly prompting ChatGPT-3.5 as both query rewriters and rewrite editors twice to incorporate all the desirable properties for producing the final rewritten queries.

\textbf{RETPO}~\cite{yoon2024ask}: A retriever preference adapted query rewriting method that fine-tunes LLaMA-2-7B as a QR model with an external QR dataset generated by GPT-4.

\textbf{ConvDR}~\cite{yu2021few}: A traditional conversational dense retrieval method that uses knowledge distillation to learn the session embeddings with relevance judgments from the human-rewritten queries based on the ANCE model.

\textbf{InstructorR}~\cite{jin2023instructor}: A LLM-based general conversational dense retriever tailored to various tasks and domains by fine-tuned with various task-specific instructions and relevance judgments based on FlanT5-XL model.

\textbf{RepLLaMA}~\cite{ma2023fine}: A large ad-hoc dense retriever fine-tuned on top of the LLaMA-7B model on the MSMARCO dataset.

\textbf{E5-Mistral}~\cite{wang2024improving}: A large ad-hoc retriever fine-tuned on top of Mistral-7B model on the synthetic dataset generated by ChatGPT-3.5 and MSMARCO.

\textbf{LLM-Embedder}~\cite{zhang2023retrieve}: A unified retrieval model that can support diverse retrieval augmentation needs of LLMs, which is fine-tuned on various tasks and datasets such as MSMARCO, NQ, ToolLLM, QReCC, FLAN, Books3, and Multi-Session Chat.

\textbf{HyDE}~\cite{gao2023precise}: A zero-shot retrieval method, which adopts ChatGPT-3.5 to generate hypothetical documents for the query,
then retrieves real documents with hypothetical documents.

\textbf{Query2doc}~\cite{wang2023query2doc}: A
zero-shot query expansion approach, which expands the original query with the generated documents from ChatGPT-3.5.

\textbf{LLM4CS}~\cite{mao2023large}: A state-of-the-art LLM-based prompting method for conversational query rewriting. We implement it with full aggregation by calling LLMs five times for the query and response generation but without the chain-of-thought (CoT) content because of the efficient annotation consideration in practical scenarios.

\section{Results of Mistral-2-7B}
\label{app:Main Results of Mistral}

\autoref{tab:main_mistral} and \autoref{tab:cast_mistral} show the performances of our methods using Mistral-2-7B as the backbone LLM, replacing LLama-2-7B as previously shown in \autoref{tab:main_res} and \autoref{tab:cast_main}, respectively. These results exhibit trends similar to those observed with LLama-2-7B, as discussed in Sections \ref{sec:Main Results} and \ref{sec:Zero-shot Results}. More specifically, \texttt{\fname{-AD}} consistently outperforms \texttt{\fname{-FT}} across all settings, and combining the results of both methods (\texttt{\fname{-Fusion}}) yields the best performance. It is worth noting that results with Mistral-2-7B are systematically lower than the ones with LLama-2-7B on most datasets (except on CAsT-21). 
Besides, by comparing the fusion performance between Mistrial and LLaMa, we notice that in most cases when the gap between \texttt{CHIQ-AD} and \texttt{CHIQ-FT} is large, \texttt{CHIQ-Fusion} results are either slightly better or worse than \texttt{CHIQ-AD}. This is mainly because the poor quality of the rank list obtained by \texttt{CHIQ-FT} negatively impacts the one from \texttt{CHIQ-AD}. However, when the gap is smaller, we notice a significant gain for \texttt{CHIQ-Fusion}, suggesting that both variants are generating good and complementary rank lists. It will be interesting to investigate how we can better take advantage of \texttt{CHIQ-AD} and \texttt{CHIQ-FT} in an adaptive fusion.
Nevertheless, performing QR on top of the enhanced history with our approach still outperforms most other settings and datasets.

\section{Case Analysis}
\label{app:Case Analysis}

\autoref{tab:case1} and \autoref{tab:case2} showcase two examples that support the case study analysis conducted in \S~\ref{sec:Case Analysis}. In the example of \autoref{tab:case1}, \textbf{QD} and  \textbf{RE} \textbf{QD} and \textbf{RE} contribute by adding the terms ``hormone'' into ``cortisol, glucagon, adrenaline, cytokines orexin, and melatonin'' to the enhanced history. In addition,   \textbf{PR} enriches the emotional context by including ``excitement, anxiety, or fear'', which co-occurs in the gold positive passage, thereby improving the scores. Since a topic switch (\textbf{TS}) is detected, earlier turns containing noisy terms, such as ``Adenosine triphosphate (ATP)'', are dropped from the history summary (\textbf{HS}). Therefore, performing QR with our methods on the enhanced history results in top passages being ranked higher compared to those based on the original history (LLM-QR). Similar trends are observed in the second case shown in Table~\ref{tab:case2}, where \texttt{\fname{-FT}} and \texttt{\fname{-AD}} outperform QR with the original history. This improvement may be attributed to the enhanced history increasing semantic similarity through references to titles of Tjader's artworks from \textbf{PR} and \textbf{HS}, such as ``Inauguration of the Pleasure Dome''. However, names of various collaborators like ``Kenneth Anger and Stan Brakhage'' may introduce noise in \textbf{PR}.

\begin{table*}[t]
    \centering
    \resizebox{\textwidth}{!}{
    \begin{tabular}{l|l}
    \toprule
    \textbf{Original History} & \textbf{Our Enhanced History} \\ 
    \midrule
    $u_1$: The primary high energy mole-  & \textbf{QD:} Which hormone among \textcolor{blue}{cortisol, glucagon, adrenaline, cytokines orexin,} \\
    cule in human metabolism is? & \textcolor{blue}{and melatonin} is associated with an emotional response? \qquad \textbf{TS}: Yes\\
    $r_1$: Adenosine Triphosphate (ATP). & \textbf{RE:} These hormones, including \textcolor{blue}{cortisol, glucagon, adrenaline, cytokines, ore-} \\
    $u_2$: What is catabolism? & \textcolor{blue}{xin, and melatonin}, play various roles in regulating \textcolor{orange}{metabolic processes} invo-\\
    $r_2$: It is the set of metabolic processes & lving the breakdown of larger molecules to \textcolor{orange}{produce ATP during catabolism}.\\    
    that breaks down large molecules. & \textbf{PR:} The emotional responses to \textcolor{blue}{hormones} can vary greatly among individuals. \\
    $u_3$: Which \underline{hormones} are related to it? & \textcolor{blue}{Adrenaline (epinephrine) is often associated with excitement, anxiety, or fear.}\\
    $r_3$: Cortisol, Glucagon, \underline{Adrenaline}, & \textbf{HS:} \textcolor{orange}{Adenosine triphosphate (ATP)} serves as the primary high energy molecule \\
    Cytokines, Orexin, and Melatonin. & in \textcolor{orange}{human metabolism. Catabolism refers to the metabolic processes} that break \\
    $u_4$: What is the emotional response & down large molecules into smaller ones. Cortisol, glucagon, adrenaline, cyto- \\
    due to the \underline{third one}? & kines, orexin, and melatonin are associated hormones involved in this process. \\
    \toprule
    \textbf{Rewritten Query} $q_4^{\prime}$ & \textbf{Gold Positive Passage} $p_4^{*}$ \\ 
    \midrule
    LLM-QR: What is the emotional & Every emotional response has a behavioral component, an autonomic compon- \\
    response associated with the third & ent, and a \textcolor{blue}{hormonal component. The hormonal component} includes the release \\
    \underline{hormone}? \textbf{(Rank: 41)}& of \textcolor{blue}{adrenaline, an adrenomedullary} response that occurs in response to stress and\\
    \fname{-AD}: QD + PR + RE \textbf{(Rank: 4)}& that is controlled by the sympathetic nervous system. The major emotion studi- \\
    \fname{-FT}: How does \underline{adrenaline} & ed in relation to \textcolor{blue}{adrenaline is fear}. In an experiment, subjects who were injected \\
    impact mood? \textbf{(Rank: 3)} & with \textcolor{blue}{adrenaline expressed more negative and fewer positive facial expressions} \\
    \fname{-Fusion} \textbf{(Rank: 1)} & to fear films compared to a control group. \\
    \bottomrule
    \end{tabular}
    }
    \caption{The first case for analyzing the content between the original and our enhanced history, as well as the rewritten queries generated on top of them. The \textcolor{blue}{blue} tokens and the \textcolor{orange}{orange} tokens stand for the effective and noisy patterns for matching, respectively. The \underline{underline} tokens denote the coreference relation between rewritten queries and the original context. The \textbf{Rank} indicates the ranking position of the gold positive passage w.r.t each query.}
    \label{tab:case1}
\end{table*}

\begin{table*}[t]
    \centering
    \resizebox{\textwidth}{!}{
    \begin{tabular}{l|l}
    \toprule
    \textbf{Original History} & \textbf{Our Enhanced History} \\ 
    \midrule
    $u_1$: Are there any interesting aspects about Cal Tjader? & \textbf{QD:} What was the outcome or reception of Kenneth Anger's collab- \\
    $r_1$: Cal Tjader entered the United States Navy in 1943 & oration with Stan Brakhage on their film project in the United States? \\
    served as a medical corpsman in the Pacific Theater & \textbf{PR:} It is unclear if \textcolor{orange}{Kenneth Anger and Stan Brakhage's} confiscated \\
    until March 1946. & film was successful as it is presumed to have been destroyed. \\
    $u_2$: Were any albums released? & However, Anger's later film, \textcolor{blue}{Inauguration of the Pleasure Dome} \\
    $r_2$: At San Francisco State he met Dave Brubeck, a & (1954), received critical acclaim for its themes. \qquad \textbf{TS:} No\\
    young pianist fresh from a stint in the Army. Brubeck & \textbf{HS:} In 1953, Kenneth Anger returned to the United States following \\
    introduced Tjader to Paul Desmond. & his mother's death to handle her estate. There, he befriended Stan \\
    $u_3$: What were the jazz album released for? & Brakhage and together they produced a controversial film, which was \\
    $r_3$: Cal Tjader continued the trio work in California & confiscated and likely destroyed due to its obscene content. Despite \\
    with bassist Jack Weeks and pianists John Marabuto & this setback, Anger went on to create his groundbreaking 38-minute \\
    ... recording his first LP with them for Fantasy Records. & surrealist work, \textcolor{blue}{"Inauguration of the Pleasure Dome,"} in 1954,\\
    $u_4$: What was a title of \underline{one of the albums}? & \textcolor{blue}{showcasing Crowleyan and Thelemite themes}.\\
    \toprule
    \textbf{Rewritten Query} $q_4^{\prime}$ & \textbf{Gold Positive Passage} $p_4^{*}$ \\ 
    \midrule
    LLM-QR: Which album did \underline{Cal Tjader record with} & The most obvious deviation from Tjader's Latin jazz sound was \\
    \underline{Jack Weeks and John Marabuto} in California?\textbf{(Rank:16)} & \textcolor{blue}{Several Shades of Jade and the follow-up Breeze From the East}. \\
    \fname{-AD}: QD + PR + HS \textbf{(Rank: 7)} & Both albums attempted to combine \textcolor{blue}{jazz and Asian music}, much as \\
    \fname{-FT}: Which specific album title did \underline{Cal Tjader} & Tjader and others had done with Afro-Cuban. The result was ...\\
    \underline{record with Jack Weeks on bass and either John} & Other experiments were not so easily dismissed. Tjader teamed up \\
    \underline{with Marabuto or Vince Guaraldi} on piano? \textbf{(Rank: 10)} & \textcolor{blue}{New Yorker Eddie Palmieri} in 1966 to produce \textcolor{blue}{El Sonido} \\
    \fname{-Fusion}: \textbf{(Rank: 5)} & \textcolor{blue}{Nuevo} A companion LP was recorded for Palmieri's ...\\
    \bottomrule
    \end{tabular}
    }
    \caption{The second case for analyzing the content between the original and our enhanced history, as well as the rewritten queries generated on top of them. The indication is consistent with Table~\ref{tab:case1}.}
    \label{tab:case2}
\end{table*}

\end{document}